\documentclass[a4paper,fleqn,usenatbib]{mnras}

\usepackage{txfonts}

\usepackage[T1]{fontenc}
\usepackage{ae,aecompl}

\usepackage[toc,page]{appendix}
\usepackage{natbib}
\usepackage{amssymb}
\usepackage{epsf}
\usepackage{pstricks}
\usepackage{color}
\usepackage{graphicx}
\usepackage{slashed}
\usepackage{relsize}
\usepackage{morefloats}
\usepackage{multirow}
\usepackage{array}
\usepackage{hyperref}
\usepackage[justification=centering]{caption}
\usepackage{url}

\title[Peculiar Spin Frequency Evolution of PSR~J1119$-$6127]{Peculiar Spin Frequency and Radio Profile Evolution of PSR~J1119$-$6127 Following Magnetar-like X-ray Bursts}

\author[S. Dai et al.]
{S. Dai$^{1}$\thanks{E-mail: shi.dai@csiro.au}, S. Johnston$^{1}$, P. Weltevrede$^{2}$, M. Kerr$^{3}$, M. Burgay$^{4}$, P. Esposito$^{5}$, 
\newauthor G. Israel$^{6}$, A. Possenti$^{4}$, N. Rea$^{5,7,8}$, J. Sarkissian$^{9}$\\
$^{1}$CSIRO Astronomy and Space Science, Australia Telescope National Facility, Box 76 Epping NSW 1710, Australia\\
$^{2}$Jodrell Bank Centre for Astrophysics, The University of Manchester, Alan Turing Building, Manchester, M13 9PL, UK\\
$^{3}$Space Science Division, Naval Research Laboratory, Washington, DC 20375-5352, USA\\
$^{4}$INAF Osservatorio Astronomico di Cagliari, Via della Scienza 5, I-09047 Selargius, Italy\\
$^{5}$Anton Pannekoek Institute for Astronomy, University of Amsterdam, Postbus 94249, 1090GE Amsterdam, The Netherlands\\
$^{6}$INAF Osservatorio Astronomico di Roma, via Frascati 33, I-00040 Monteporzio Catone, Roma, Italy\\
$^{7}$Institute of Space Sciences (ICE, CSIC), Campus UAB, Carrer de Can Magrans s/n, 08193 Barcelona, Spain\\
$^{8}$Institut d'Estudis Espacials de Catalunya (IEEC), 08034 Barcelona, Spain\\
$^{9}$CSIRO Astronomy and Space Science, Parkes Observatory, PO Box 276, Parkes NSW 2870, Australia\\
}

\date{Accepted XXX. Received YYY; in original form ZZZ}

\pubyear{2017}

\begin{document}
\label{firstpage}
\pagerange{\pageref{firstpage}--\pageref{lastpage}}
\maketitle

\begin{abstract}
We present the spin frequency and profile evolution of the radio pulsar J1119$-$6127 following magnetar-like X-ray 
bursts from the system in 2016 July. Using data from the Parkes radio telescope, we observe 
a smooth and fast spin-down process subsequent to the X-ray bursts resulting in a net change in 
the pulsar rotational frequency of $\Delta\nu\approx-4\times10^{-4}$\,Hz.
During the transition, a net spin-down rate increase of 
$\Delta\dot\nu\approx-1\times10^{-10}$\,Hz\,s$^{-1}$ is observed, followed
by a return of $\dot{\nu}$ to its original value.
In addition, the radio pulsations disappeared after the X-ray bursts and 
reappeared about two weeks later with the flux density at 1.4\,GHz 
increased by a factor of five. The flux density then 
decreased and undershot the normal flux density followed by a slow recovery back to normal.
The pulsar's integrated profile underwent dramatic and short-term
changes in total intensity, polarization and position angle.
Despite the complex evolution, we observe correlations between the spin-down 
rate, pulse profile shape and radio flux density.
Strong single pulses have been detected after the X-ray bursts with their 
energy distributions evolving with time.
The peculiar but smooth spin frequency evolution of PSR~J1119$-$6127 accompanied by 
systematic pulse profile and flux density changes are most likely to 
be a result of either reconfiguration of the surface magnetic fields or particle winds 
triggered by the X-ray bursts.
The recovery of spin-down rate and pulse profile to normal provides us the best case 
to study the connection between high magnetic-field pulsars and magnetars.

\end{abstract}

\begin{keywords}
pulsars: general
\end{keywords}



\section{Introduction} \label{sec:intro}

Of the more than 2600 pulsars now known, the so-called magnetars form
a small sub-class with only 29 members\footnote{See http://www.physics.mcgill.ca/~pulsar/magnetar/main.html~\citep{ok14}}~\citep[see e.g.,][for reviews]{kb17,eri18}.
Typically, these pulsars have long rotation periods and very high spin-down rates, hence high
inferred surface magnetic fields above a few $10^{13}$\,G.
Magnetars are characterised by X-ray outbursts whose luminosity exceeds the 
spin-down luminosity and are therefore likely to be powered by the ultra-high 
magnetic fields. 
In addition, some magnetars have been detected as transient pulsating radio sources, 
showing several differences from the normal radio pulsars, including extreme variability 
in their flux densities, pulse profiles and spin-down rates~\citep[e.g.,][]{crh+06,crj+07,ccr+07,crj+08,lbb+12}.

If energetic burst activities of magnetars are powered by their magnetic fields, 
and given that similar behaviour has been seen in sources with inferred magnetic fields 
in the range of ordinary radio pulsars~\citep[such as SGR 0418+5729 and Swift J1822.3$-$1606;][]{ret+10,rie+12}, 
it is then conceivable that radio pulsars with high inferred surface dipole magnetic 
fields (hereafter high-$B$ pulsars) could exhibit magnetar-like activity~\citep{km05,nk11}.
Studies have also been carried out to unify the observational diversity of
magnetars, high-B radio pulsars, and isolated nearby neutron
stars via magneto-thermal evolution models~\citep[e.g.,][]{vrp+13}. 
Previously, only one high-$B$, rotation-powered pulsar, PSR~J1846$-$0258, 
has been observed to transition to a magnetar-like phase~\citep{ggg+08}. While 
several short magnetar-like bursts and a sudden spin-up glitch followed by 
a strong over-recovery were detected~\citep{kh09,lkg10,lnk+11}, no radio 
activities have been observed.
More observational evidences of high-$B$ radio pulsars in the transitional phase 
are therefore important for us to understand their connection to magnetars, 
and to unify these two populations of isolated neutron stars.

PSR~J1119$-$6127 is a high-$B$, rotation-powered radio pulsar
discovered in the Parkes multibeam pulsar survey~\citep{ckl+00}; the
pulsar's basic timing parameters are listed in Table~\ref{param}.
It has been regularly observed at the Parkes radio telescope as a part of a 
monitoring program of energetic pulsars~\citep{wjm+10,aaa+13} and two 
unusual glitches have been detected in the pulsar described in full by \citet{wje11} and \citet{awe+15}.
In addition, \citet{wje11} also detected extremely rare intermittent 
pulse components following the glitch.
\citet{awe+15} measured a braking index of $\simeq2.7$, 
indicating that in the $P-\dot{P}$ plane the pulsar appears to be moving 
towards the magnetar population. They also found marginal evidence for a 
permanent change in the braking index following a glitch.

Magnetar-like X-ray bursts from PSR~J1119$-$6127 were
detected by the \textit{Fermi}/Gamma-ray Burst Monitor on 2016 July 27~\citep{ykr16} and 
by the \textit{Swift}/Burst Alert Telescope on July 28~\citep{klm+16}. This 
made PSR~J1119$-$6127 the first rotation-powered, radio pulsar to show magnetar-like activities.
Following the X-ray bursts, the persistent X-ray flux of PSR~J1119$-$6127 
increased by a factor of 160, and the pulsar underwent a large glitch~\citep{akt+16}. 
Extensive searches in both Fermi and Swift data for lower luminosity bursts 
uncovered 10 additional bursts from the source~\citep{glk+16}. 
Additionally, its pulsed radio emission at $\sim1.4$\,GHz was quenched following the bursts~\citep{bpk+16a}, 
and only reappeared some two weeks later~\citep{bpk+16b}. Higher radio frequency observations 
confirmed the initial disappearance of the radio emission and its prompt reactivation 
and also showed pulse profile evolution over several months of observation~\citep{mpd+17}.
Simultaneous observations at X-ray, with XMM-Newton and NuSTAR, and at radio frequencies
with the Parkes radio telescope, showed that the 
rotationally powered radio emission shut off multiple times, in coincident with the 
occurrence of X-ray bursts~\citep{abl+17}.
Three months after the bursts, the pulsar was still brighter in the X-rays by a factor of 22 in 
comparison with its quiescence and the X-ray images revealed a nebula brighter than 
in the pre-burst Chandra observations~\citep{bsm17}.

In this paper, we present observations of the spin frequency and radio profile evolution of PSR~J1119$-$6127 
after the X-ray outburst using data taken with the Parkes radio telescope. 
The peculiar spin frequency evolution of PSR~J1119$-$6127 together 
with its magnetar-like activities, provide us an opportunity to study the relation/transition 
between high-$B$ pulsars and magnetars and to understand the origin of X-ray bursts and 
the evolution of magnetic fields.
In Section~\ref{sec:observations} we describe the observations and the data reduction.
In Section~\ref{sec:result} we present the results on spin frequency evolution, flux density, 
polarization pulse profiles and single pulses. Discussions and a summary are given in 
Section~\ref{sec:discussion} and \ref{sec:sum}.

\section{Observations and Data Reduction}
\label{sec:observations}
PSR~J1119$-$6127 is observed regularly using the Parkes radio telescope
under the auspices of program P574, with an approximate observing cadence of four weeks. 
Data taken prior to 2014 are described in \citet{awe+15}. In this paper we use observations 
taken since 2015\footnote{When investigating the spin frequency evolution we included spin
frequency measurements back to 2013.}, which were primarily carried out in the band 
centred at 1369\,MHz using the central beam of the 20\,cm multibeam receiver.
From February to October 2016 the H-OH receiver centred at 1465\,MHz was used.

Triggered by the X-ray bursts on 2016 July 27, PSR~J1119$-$6127 was extensively
observed in 2016 July, August and September using the Parkes radio telescope
as a part of proposal P626 (PI: M. Burgay). 
Two observing modes, fold-mode and search-mode, were used for the follow-up observations. 
For fold-mode, data were folded modulo the pulse period with 1024 phase bins in each of 1024 
frequency channels, integrated for 30\,s and written to disk. 
For search-mode, data were recorded with 4-bit-sampling every 256\,$\mu$s 
in each of 512 frequency channels. For both modes, full Stokes information was recorded.
The observing frequency, bandwidth and integration time are presented in Table~\ref{num_pulse}. 
\begin{table}
\begin{center}
\caption{Spin parameters for PSR~J1119$-$6127 following the initial X-ray burst. Error bars on the last digit are given in parentheses.} 
\label{param}
\begin{tabular}{lc}
\hline
\hline
	Epoch (MJD)       &   57935.25    \\
	Spin frequency ($\nu$) & 2.438745737(4)\,Hz\\ 
	Spin-frequency derivative ($\dot{\nu}$) & $-2.466288(9)\times10^{-11}$\,Hz\,s$^{-1}$ \\ 
	Spin-frequency second derivative ($\ddot{\nu}$) & $3.80(1)\times10^{-21}$\,Hz\,s$^{-2}$ \\ 
	MJD range         & 57795$-$58208  \\
\hline
\end{tabular}
\end{center}
\end{table}

For the fold-mode observations, the data processing procedure follows 
that described in \citet{kcw+18}. We removed 5 per cent of the bandpass at
each edge to mitigate RFI and aliasing and excised data affected by narrow-band and impulsive
radio-frequency interference for each sub-integration. 
To measure the differential gains between the signal paths of the
two voltage probes, once an hour we observe a pulsed noise signal
injected into the signal path prior to the first-stage low-noise amplifiers.
The noise signal also provides a reference brightness for each
observation. To correct for the absolute gain of the system, we make
use of observations of the radio galaxy 3C 218 (Hydra A); by using
on- and off-source pointings, we can measure the apparent brightness
of the noise diode as a function of radio frequency. 
A dispersion measure (DM) of 704.8\,cm$^{-3}$\,pc~\citep{pkj+13} and a rotation measure 
(RM) of 853\,rad\,m$^{-2}$~\citep{jw06} were used in the data processing.
All data reduction made use of the {\sc Psrchive} \citep{hvm04} software package.

\section{Results}
\label{sec:result}

\subsection{Spin frequency evolution}
\label{sec:nu}

There are a number of difficulties in solving for the evolution of the
pulsar's spin parameters after the X-ray bursts. Firstly, the profile underwent
changes (see Section~\ref{sec:prof}) for which no absolute reference point can be 
used to determine pulse time of arrivals. Secondly, standard pulsar timing software does not cope well
with the strong and rapid evolution of both $\nu$ and $\dot{\nu}$ seen here.
We have therefore not attempted a phase-coherent fit over this time interval.
Rather, we form an analytic template for each observation 
and then measure $\nu$ by fitting pulse time of arrivals of individual 
sub-integrations across the observation duration. 

In the top panel of Fig.~\ref{f0}, we show $\nu$ as a function of 
time with the $\dot{\nu}$ measured by \citet{awe+15} shown as a red line and 
the X-ray outburst shown as the yellow region.
Before the X-ray bursts, $\dot{\nu}$ agrees with the previous measurement.
However, after the X-ray bursts, $\nu$
starts to decrease rapidly as shown in the second 
panel of Fig.~\ref{f0} (created after removing the slope due to $\dot{\nu}$).
\citet{akt+16} reported a large glitch at $\rm{MJD}=57596.547$ ($\Delta\nu=1.42(2)\times10^{-5}$\,s$^{-1}$), 
but because of the absence of radio emission during that period of time, we cannot 
confirm the glitch and did not observe significant discrete changes of $\nu$.
On the contrary, after the reactivation of radio emission we observed a
relatively fast and smooth decrease in $\nu$ followed 
by a rapid drop. This is completely at odds with normal glitches, which
have the opposite sign change of $\nu$ and happen virtually instantaneously~\citep[e.g.,][]{pdh+18}.
Within a time period of $\sim150$\,days, we observed a net change
in the rotational frequency by $\Delta\nu\approx-4\times10^{-4}$\,Hz.

The strong evolution of $\nu$ is accompanied by rapid changes in $\dot{\nu}$.
We fitted for $\dot{\nu}$ by using adjacent measurements of $\nu$; during 
the rapid changes of $\nu$ (MJD between 57600 and 57950), we used a sliding window 
of 10 days; for the rest of the data set, we used a sliding window of 60 days\footnote{We treated 
$\nu$ measurements before and after the X-ray bursts separately, and therefore 
$\dot{\nu}$ before and after the X-ray bursts are independent.}. 
No significant discrete change of $\dot{\nu}$ was observed after the reactivation 
of radio emission. The evolution of $\dot{\nu}$ started with a fast and
smooth decrease, peaked at $\rm{MJD}\approx57632$, and then gradually recovered 
to its original value.
Again, this is different from the recovery of $\dot{\nu}$ 
following normal glitches and previous glitches observed for this pulsar.
What is also interesting is the wiggle during the fast decrease of $\dot\nu$ (becoming more negative), 
and we will discuss this in Section~\ref{sec:discussion}.

A large spin-up glitch has been detected for PSR~J1846$-$0258 after the onset of magnetar-like
behaviour, followed by an unusually large glitch recovery~\citep{lkg10}. While the net spin-down 
of PSR~J1846$-$0258 after the event ($\Delta\nu\approx-10^{-4}$) is of the same order as we 
observed for PSR~J1119$-$6127, the evolution of $\nu$ is different.  
We did not see sudden changes in the $\nu$ and $\dot{\nu}$ for PSR~J1119$-$6127, 
and $\dot{\nu}$ peaks about 30\,days after the X-ray bursts
before recovering back to normal about 200\,days later.

An updated timing solution for the pulsar was obtained after the recovery to the
normal state ($\rm{MJD}\approx57795$), these parameters are listed in 
Table~\ref{param}.

\subsection{Flux density}
\label{sec:flux}

\begin{figure*}
\begin{center}
\includegraphics[width=7.3 in]{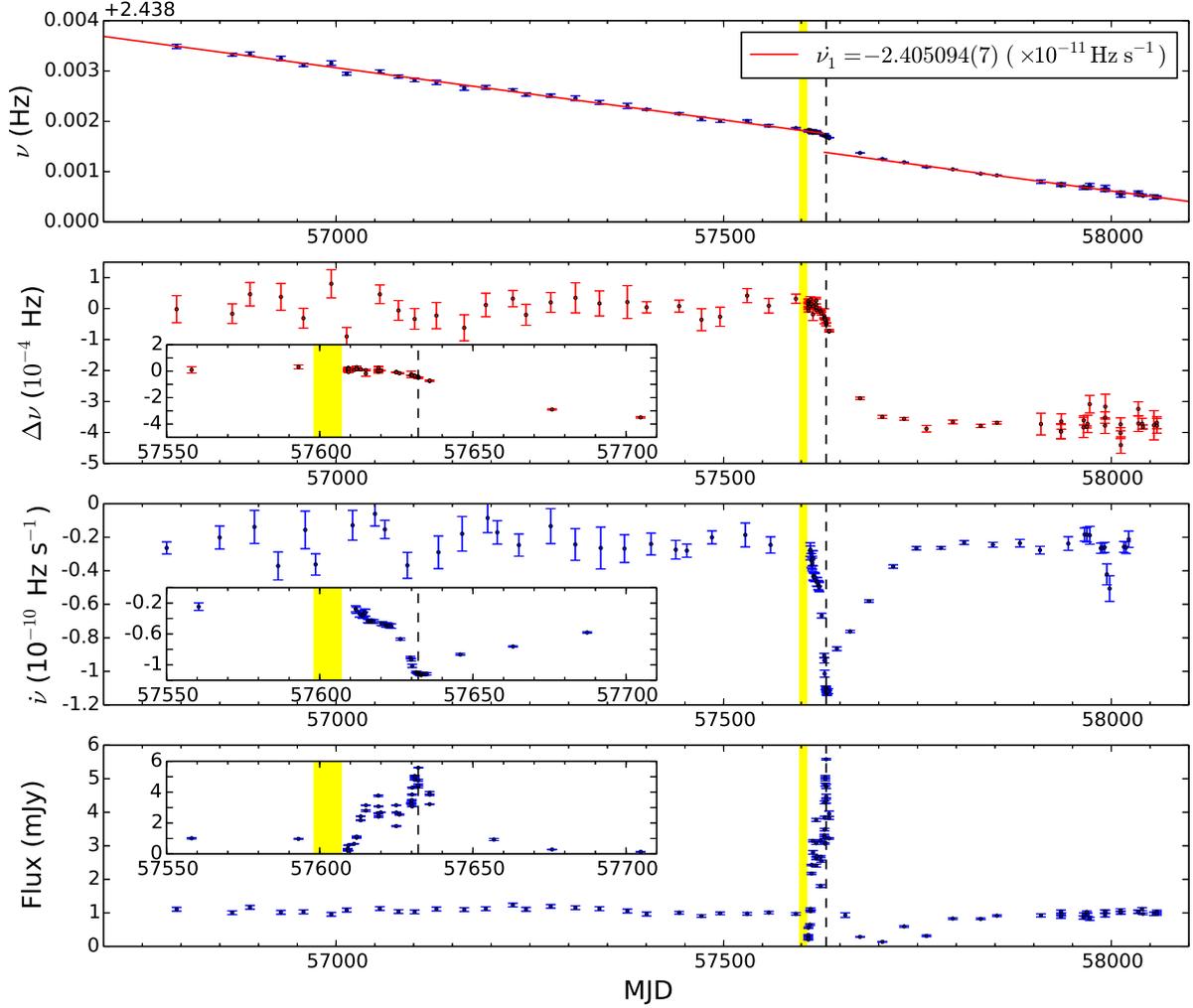}
\caption{Pulsar spin frequency ($\nu$), spin-frequency derivative ($\dot{\nu}$) and flux density 
as a function of time. Yellow regions show the radio-quiet period after the X-ray burst. Black dashed 
lines indicate the peak of flux density.}
\label{f0}
\end{center}
\end{figure*}

%
To measure the flux density, we first formed noise-free standard templates for each observation by 
fitting scaled von Mises functions (using the {\sc Psrchive} program {\sc paas}) to 
the observed profile after integrating the data over the observing band and observation duration.
The {\sc Psrchive} program {\sc psrflux} was used to measure the flux density for each observation,
which cross-correlates the observed profile with the standard template to obtain the scaling factor 
and then the averaged flux density.
The uncertainty of flux density was estimated using the standard deviation of the baseline 
fluctuations.
In the bottom panel of Fig.~\ref{f0}, we present the measured flux density as a function of time. 
After the reactivation of radio emission, the flux density increased rapidly and peaked at 
$\rm{MJD}\approx57632$ (black dashed lines) and then started decreasing. 
Intriguingly, the flux density undershot the normal flux density ($\sim1$\,mJy) and dropped to a 
minimum value of $\sim0.14$\,mJy and then slowly recovered back to normal. 
We note that the peak of flux density roughly lines up with the peak of net spin-frequency 
derivative. 
Potential correlation between reduced flux density and increased $\dot\nu$ has been observed 
in PSR J1622$-$4950~\citep{lbb+12}, but it is very different compared with what we have 
observed for PSR J1119$-$6127.
More similarly, \citet{ccr+07} reported a decrease of flux density and a broadening
of the pulse profile components accompanied by a decrease in torque of about 10\% over an 
interval of two weeks for XTE J1810$-$197. 

\subsection{Profile variations}
\label{sec:prof}
The polarization characteristics of the pulsar were described in \citet{jw06}
and in more detail by \citet{rwj17a}. In brief, in what we will refer to as 
the normal state (Profile A), the pulsar shows a single, highly linearly 
polarized component with a half-width of 19\degr~\citep{jk18} and a small amount
of circular polarization. The position angle (PA) has a shallow negative slope
as a function of phase. In the single abnormal observation directly after a 
large glitch described by \citet{wje11}, an extra component, also highly linearly polarized, appeared to lead
the normal component. \citet{rwj17a} considered both these profiles and
concluded the angle between the magnetic and rotation axes, $\alpha$, was
relatively small.

\begin{figure*}
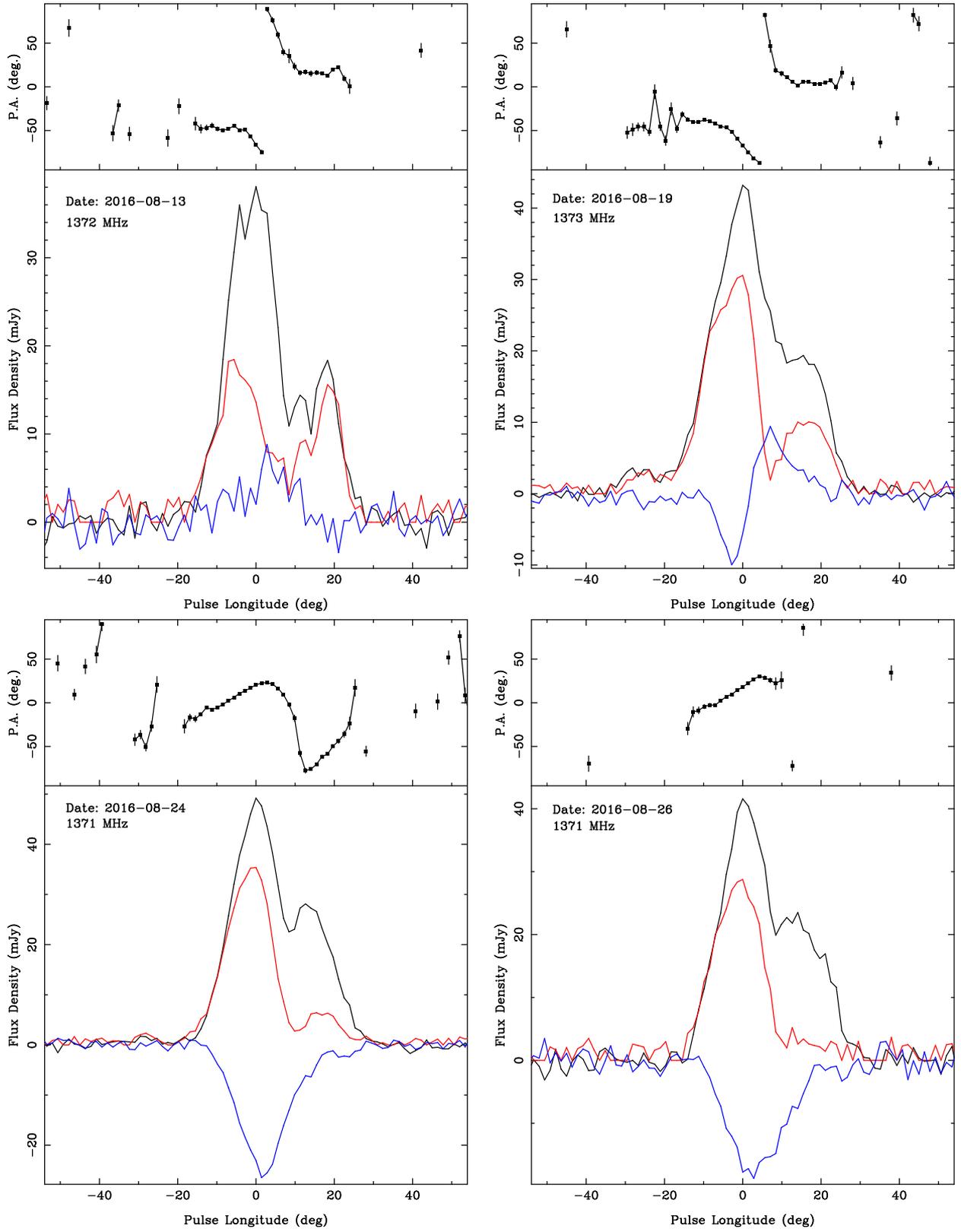

\begin{center}
\begin{tabular}{cc}
\includegraphics[width=8cm,angle=0]{type1a.ps} &
\includegraphics[width=8cm,angle=0]{type1b.ps} \\
\includegraphics[width=8cm,angle=0]{type1c.ps} &
\includegraphics[width=8cm,angle=0]{type1d.ps} \\
\end{tabular}
\end{center}
\caption{Four examples of Profile B, the two-component profile 
of PSR~J1119$-$6127.
In each panel, the black line shows total intensity, red linear polarization 
and blue circular polarization. The position angle of the linear 
polarization, referred to infinite frequency, is also shown. Phase zero
has been set according to the discussion in the text.}
\label{type1}
\end{figure*}

\begin{figure*}
\begin{center}
\begin{tabular}{cc}
\includegraphics[width=8cm,angle=0]{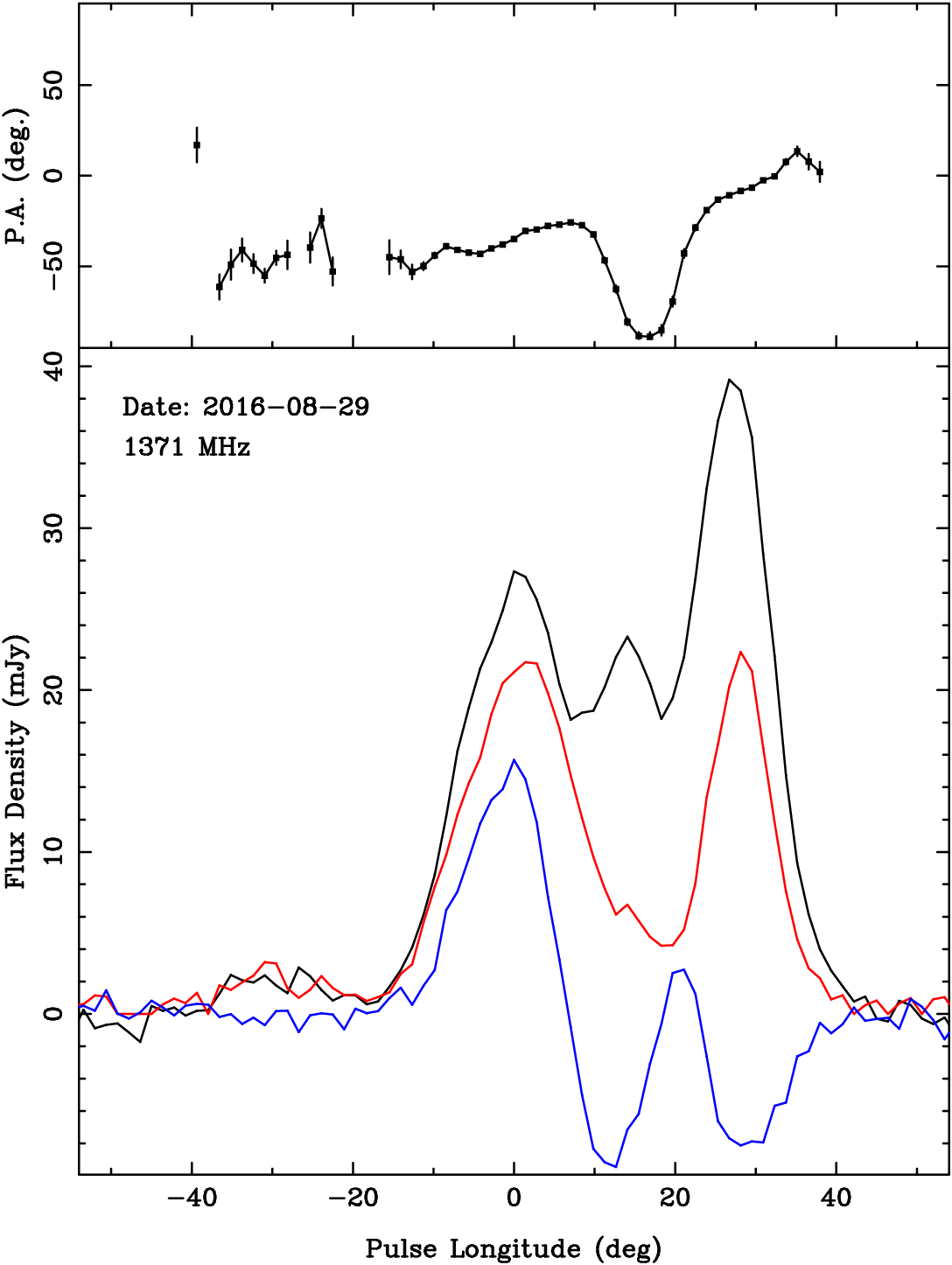} &
\includegraphics[width=8cm,angle=0]{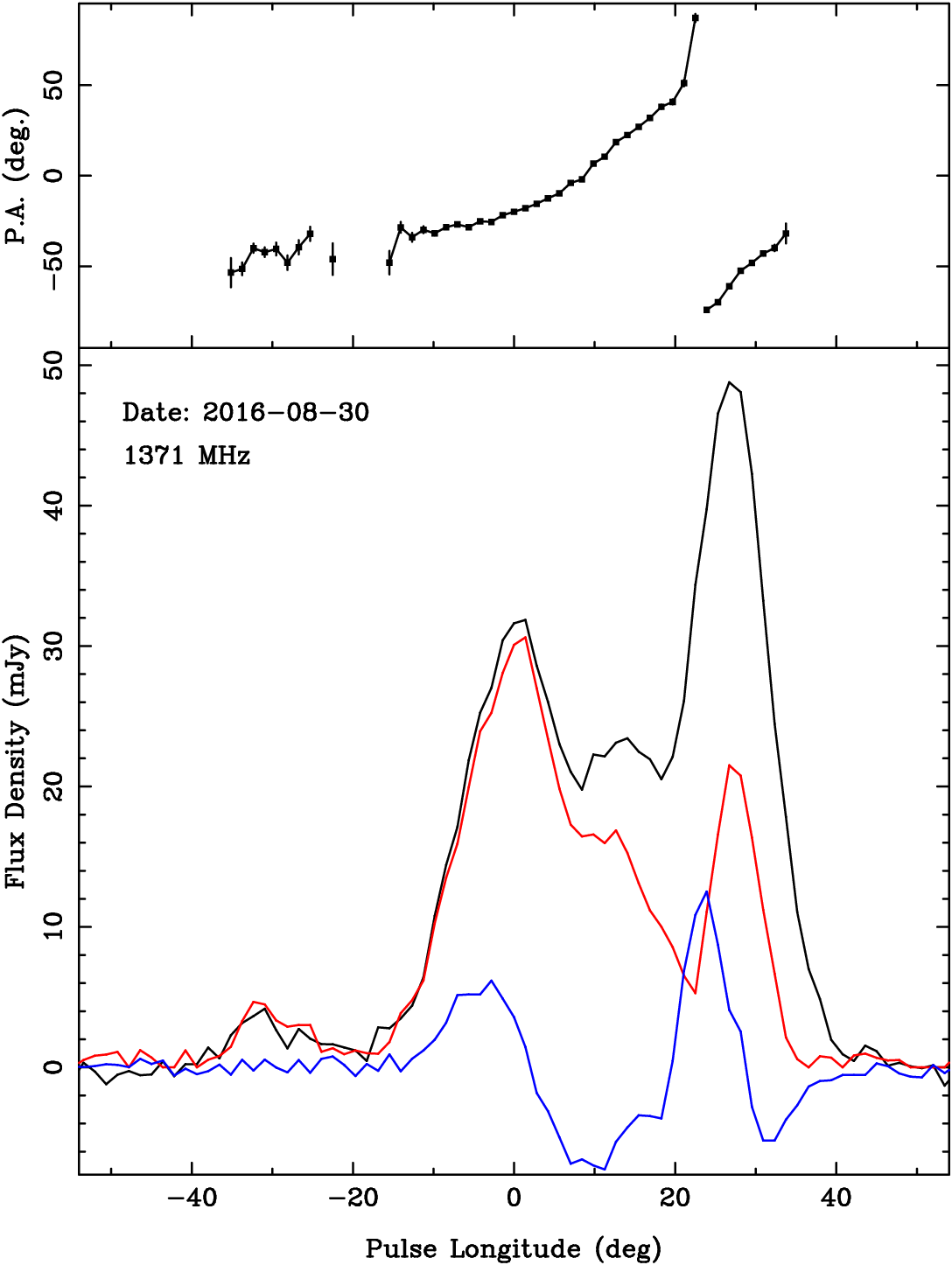} \\
\includegraphics[width=8cm,angle=0]{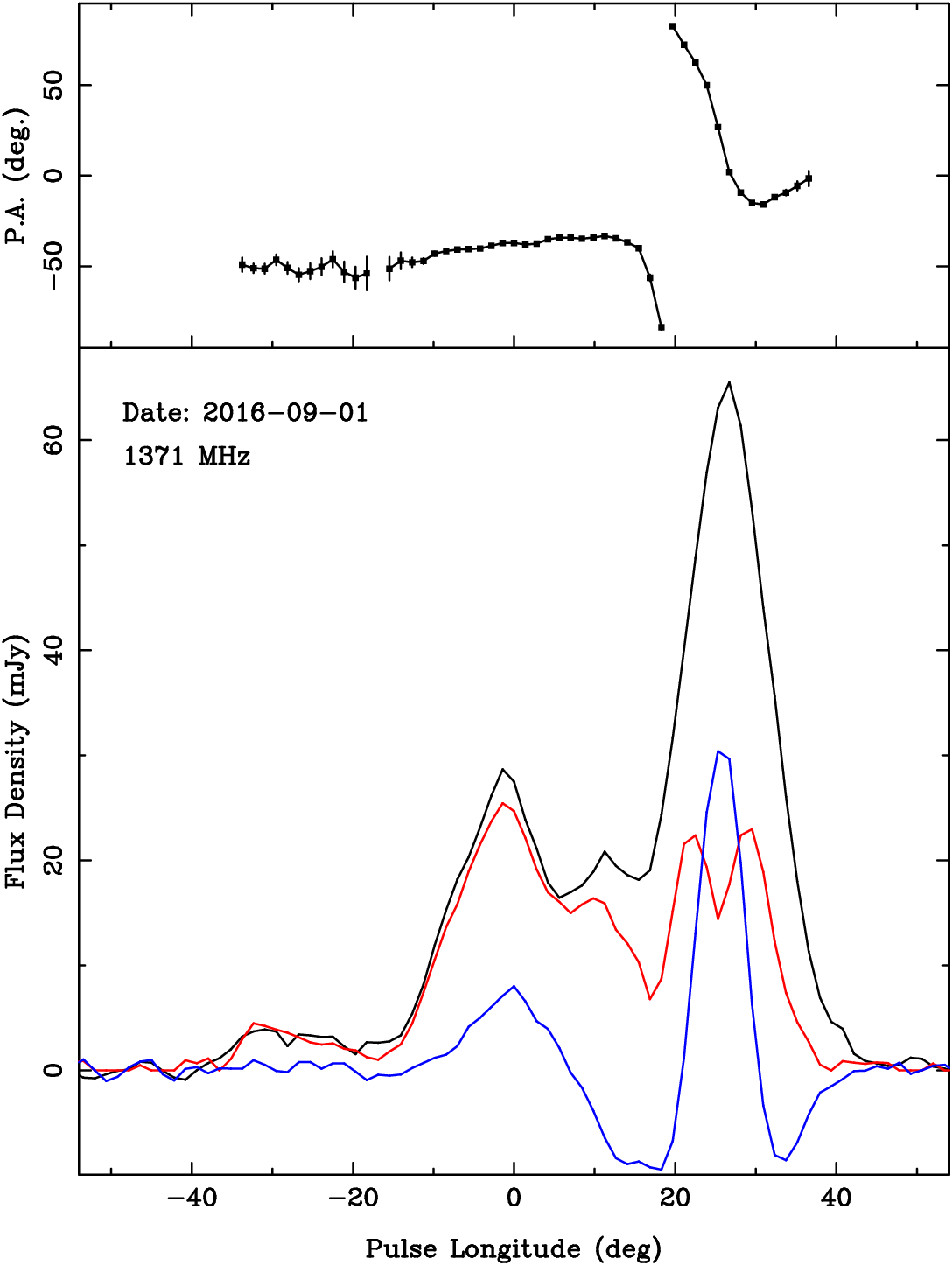} &
\includegraphics[width=8cm,angle=0]{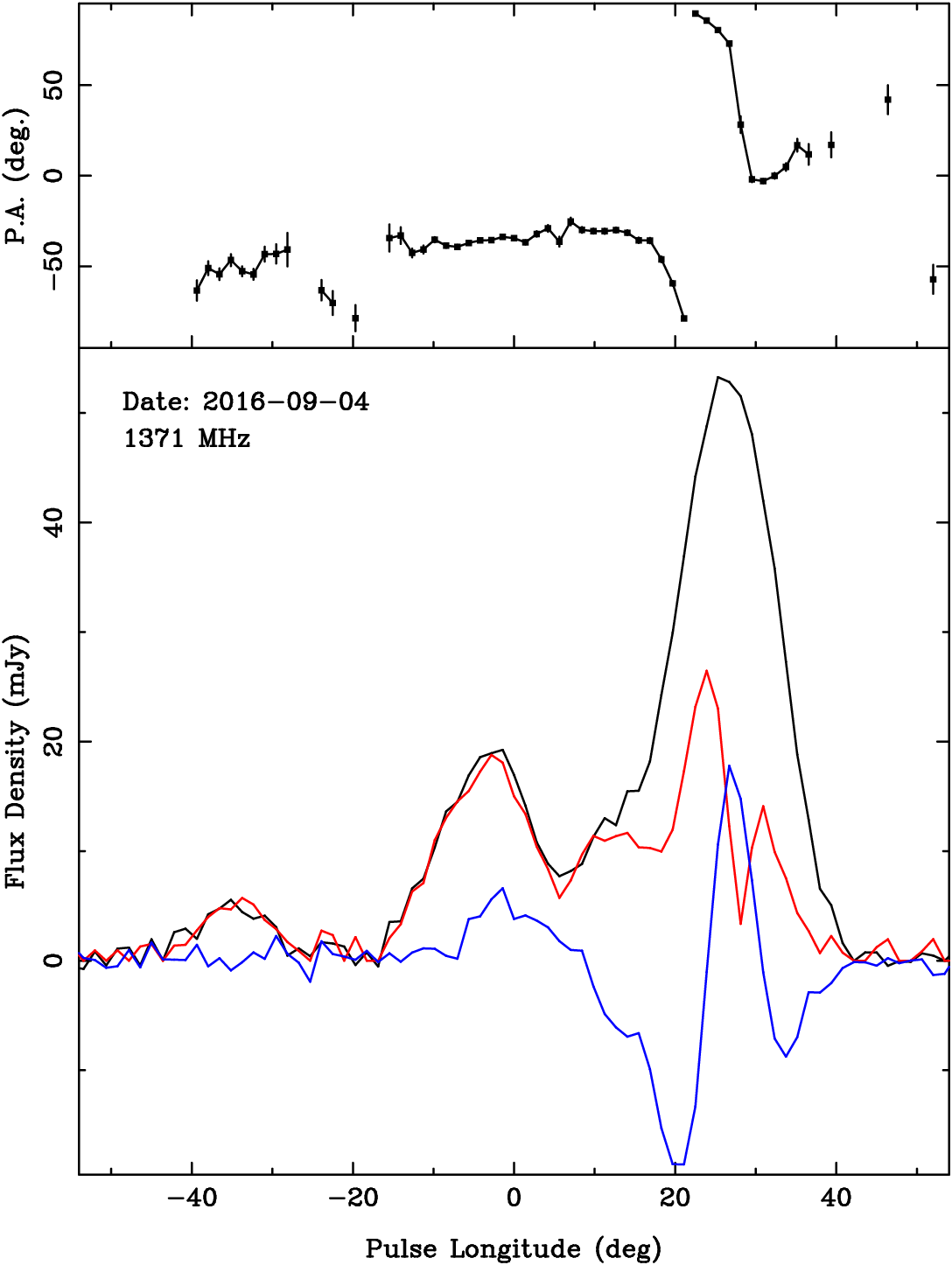} \\
\end{tabular}
\end{center}
\caption{Four examples of Profile C, the four-component profile of 
PSR~J1119$-$6127.
In each panel, the black line shows total intensity, red linear polarization 
and blue circular polarization. The position angle of the linear 
polarization, referred to infinite frequency, is also shown. Phase zero
has been set according to the discussion in the text.}
\label{type2}
\end{figure*}

In the observations under consideration here, dramatic changes in the 
pulse profile and the polarization characteristics were observed between
2016 August 12 and 2016 September 4. Broadly speaking, two alternative
profiles are observed. The first, present between 2016 August 12 and
2016 August 26 shows the presence of two components (see Figure~\ref{type1})
which we call Profile B.
The first component is highly polarized whereas the second is not.
The circular polarization underwent strong transformation within this
period as did the PA swing. In particular, the slope of the PA swing
changed from negative to positive during this timeframe. The second
state occurred between 2016 August 29 and (at least) 2016 September 4.
In this state, Profile C, there are 4 components (see Figure~\ref{type2}), 
a leading small highly polarized component separated in phase from a triple structure
with a highly polarized first component followed by two, less
polarized components. The relative amplitude of the components changes with time. The
circular polarization and the PA swing are again highly variable. As shown
above, the pulsar was much brighter during this period, largely due to the
third component of the triple structure.
There are no observations between 2016 September 4 and September 25 where the
pulsar flux density dropped from $\sim4$\,mJy to $\sim1$\,mJy, but the pulse 
profile appears to have returned to its normal profile on September 25.

Several questions arise: how do the profiles align in phase, how do the 
profiles changes correlate with the changes in $\nu$ and $\dot{\nu}$, 
and what are the causes of the profile changes?

Considering first the profile alignment, we believe the most plausible
solution is that shown in Figures~\ref{type1} and \ref{type2}, with zero phase defined as 
the peak of the profile A, the leading component of the profile B
and the leading component of profile C as shown. The
implication of this is that the normal component remains roughly
constant in amplitude through the other changes, and that the trailing, very bright
component seen at later times appears to emerge rapidly between
August 26 and 29. In addition, this would imply that the extra component
seen in \citet{wje11} is aligned in phase with the trailing bright
component seen at later times here.
What is most unexpected about these observations is the extreme
changes seen in the PA swing as a function of epoch. No evidence of changes
in PA swing has been observed before~\citep{wje11}. For many pulsars,
and for this pulsar in particular, the swing of PA is used to determine
the geometry of the star via the rotating vector model (RVM)\citep[e.g.,][]{rwj17a}.
The sign of the slope of the PA swing is used to determine whether
the line of sight cuts pole-ward or equator-ward from the magnetic axis
\citep{ew01} and therefore should not change.
The PA swing of these observations significantly deviates from 
RVM-like behaviour, and for the scope of this paper we have not attempted to 
model the swing of PA. We also note that no significant changes of RM 
have been observed following the X-ray bursts.


\subsection{Single pulses}
\label{sec:pulse}

Bright individual pulses have been observed for J1119$-$6127 previously. \citet{wje11} detected 
four strong single pulses at 1.4\,GHz on 2007 August 20 and a number of bright 
single pulses at 3.1\,GHz on 2007 July 23. The fact that the strong pulses do
not align with the peak of the averaged pulse profile suggests their RRAT-like origins.
Intriguingly, a large amplitude glitch occurred directly before these very rare  
RRAT-like events were observed. This was the first time that a glitch, or the post-glitch 
recovery process, has been observed to influence the radio emission process of a normal 
(non-RRAT) pulsar. 
A starquake model with crustal plate movement has been proposed to explain the changing 
emission properties coincident with the glitch~\citep{ags+15}, as well as the unusual 
glitch recovery~\citep[see][]{awe+15}.

\begin{figure*}
\begin{center}
\includegraphics[width=7.0 in]{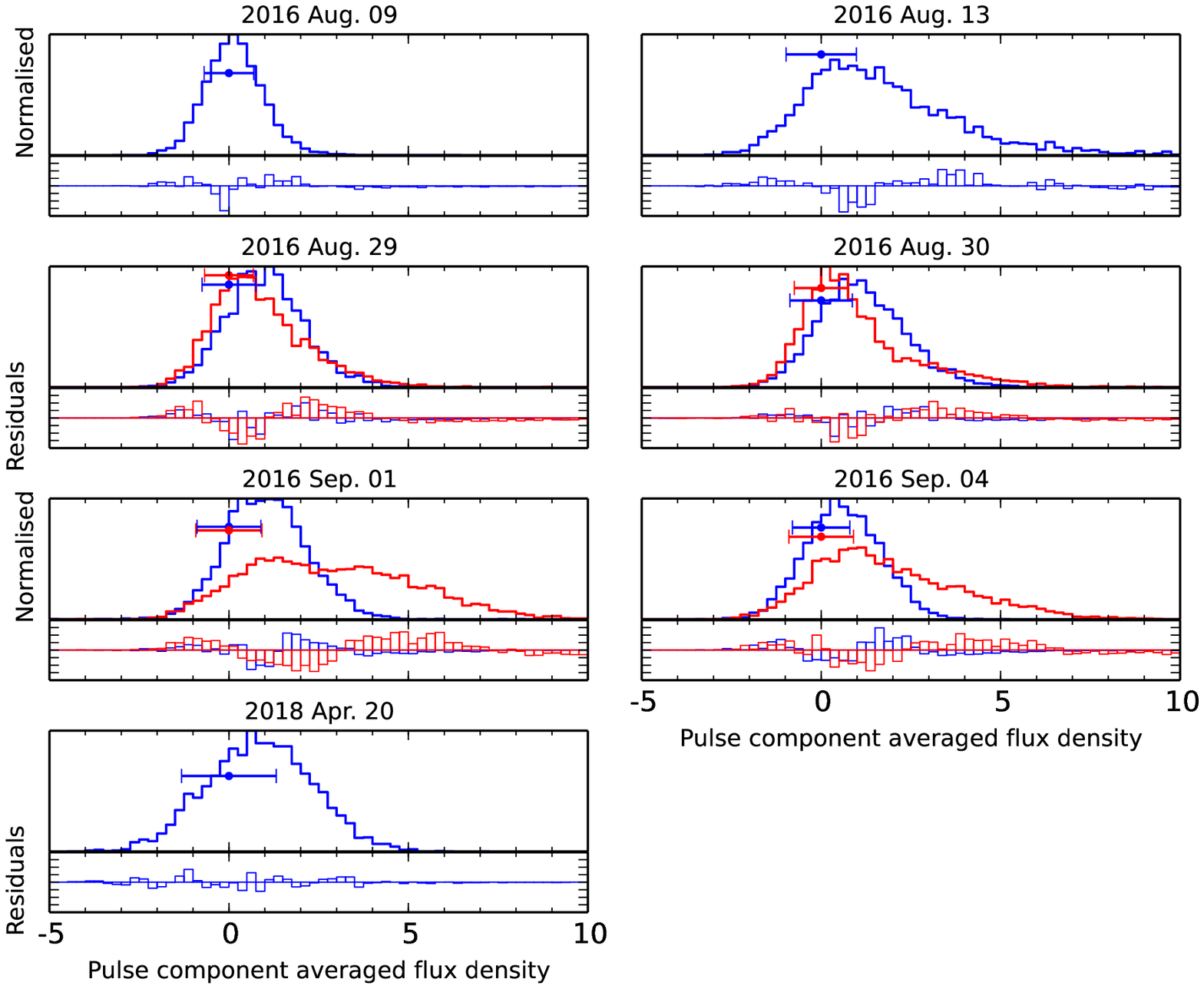}
\caption{Top panel of each subplot: averaged flux density distributions of individual pulse  
components (Blue: the main pulse component centred at a pulse longitude of zero degrees in Fig~\ref{type1} and \ref{type2}; 
Red: the trailing component centred at a pulse longitude of $\sim25$ degrees in Fig~\ref{type2}). The standard 
deviation of the noise distribution is shown as a point with horizontal error bar for each pulse component.
Bottom panel of each subplot: fitting residuals of the flux density distribution against a log-normal 
distribution convolved with the observed noise distribution. }
\label{energy}
\end{center}
\end{figure*}

After the X-ray bursts of J1119$-$6127, we also detected strong single pulses.
Out of ten search-mode observations, we detected strong single pulses with signal-to-noise (S/N) 
larger than ten in seven observations. 
In Table~\ref{num_pulse}, we list the number of detections of strong single pulses with 
$\rm{S/N}>10$ for each observation. 
To better understand the occurrence of single pulses, we produced pulse component averaged 
flux density distributions of these observations. The {\sc psrsalsa} software package~\citep{w16} 
was used to carry out manual RFI mitigation, to calculate pulse energies, 
and to fit energy distributions. In the top panel of each subplot in Fig.~\ref{energy}, we 
show pulse component averaged flux density distributions of observations taken on 2016 August 
9, 13 and 30 and on September 1 and 4, all of which show strong single pulses and have integration 
time long enough for us to study their statistics. In comparison, results of an observation 
taken during the pulsar's normal state (2018 April 20) are also shown. Flux density distributions 
of the main pulse component (centred at a pulse longitude of zero degrees in Figs~\ref{type1} and \ref{type2}) 
and the bright trailing component (centred at a pulse longitude of $\sim25$ degrees 
in Fig~\ref{type2}) are shown as blue and red histogram, respectively. The standard deviation of 
the noise distribution is shown as a point with horizontal error bar for each pulse component.
The observed energy distribution of single pulses is the intrinsic distribution 
convolved with the noise, and we fitted for the intrinsic energy distribution 
as described in \citet{w16}, assuming a log-normal distribution. In the bottom panel of 
each subplot in Fig.~\ref{energy}, we show the fitting residuals for each pulse component. 

The energy distribution of both the main and the trailing component significantly deviate
from a log-normal distribution after the X-ray bursts. For the trailing component, we 
observed a long high energy tail, and it became more and more significant as the pulsar 
became brighter and showed a new high energy component on 2016 September 1 before it faded 
away. For the main component, the energy distributions of observations with Profile C 
look similar and can be generally fitted by a log-normal distribution, but are clearly 
different from that of Profile B on 2016 August 13 which deviates from a log-normal distribution. 
In comparison, for the normal state (e.g., 2018 April 20), the intrinsic energy distribution 
of single pulses can be well described by a log-normal distribution. 
Although detailed modelling of the energy distributions is beyond the scope of this paper, 
our results show that a significantly greater number of energetic pulses were produced after 
the X-ray bursts and this can not be simply explained by the increase of the averaged flux density.
The different pulse energy distributions of two pulse profile components at different epochs 
provide evidences of changes in the pulsar magnetosphere triggered by the X-ray bursts.

\begin{figure*}
\begin{center}
\includegraphics[width=5.0 in]{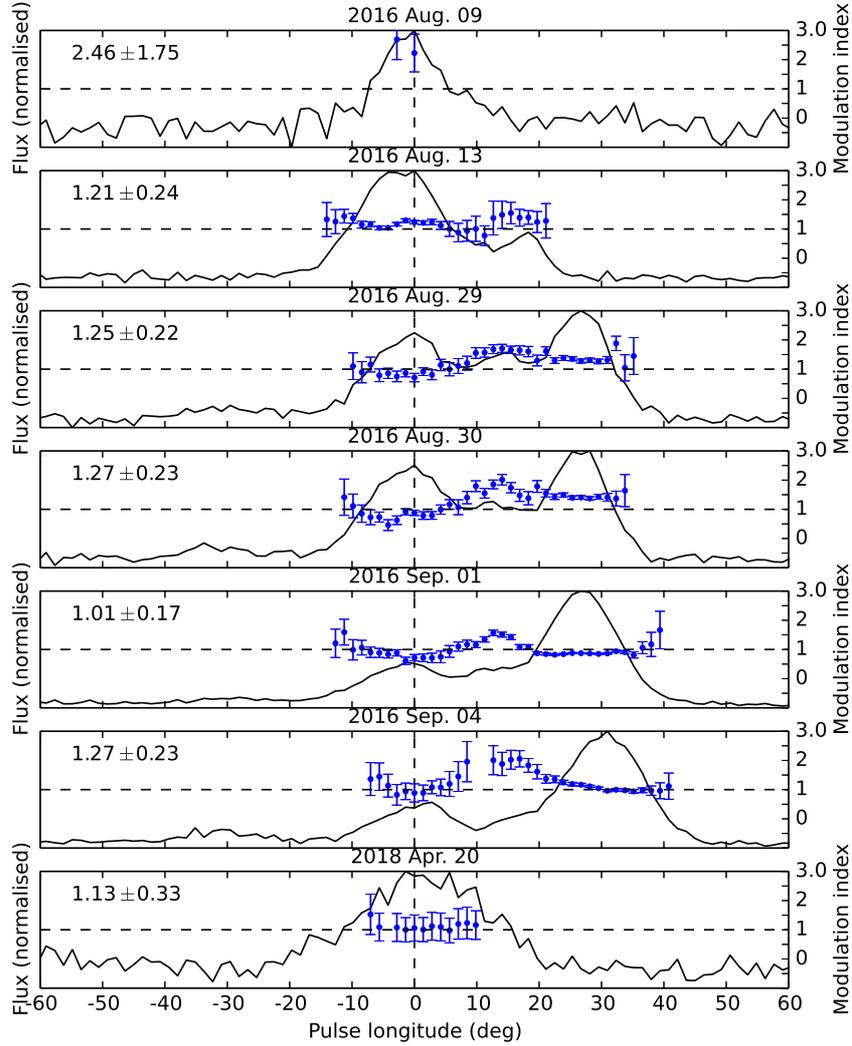}
\caption{Modulation index (blue points with error bars) as a function of pulse longitude. 
The averaged pulse profiles are shown as black lines. Averaged modulation indices are presented in
the upper left corner of each panel.}
\label{modu}
\end{center}
\end{figure*}

In order to study the variation of intensity from pulse to pulse as a function of 
pulse longitude, we calculated the modulation index defined as Eq.~1 of \citet{wje11} in the spectral domain.
In Fig.~\ref{modu}, modulation indices at different pulse phases
are shown as blue points with error bars on top of the averaged pulse profiles. 
The variation of modulation index with pulse longitude shows similar shape for all 
observations and supports our pulse profile alignment. Averaging over the on-pulse phases, 
we obtained modulation indices after the X-ray bursts (shown in the upper left corner of each panel 
of Fig.~\ref{modu}) significantly higher than those observed for most pulsars~\citep[e.g.,][]{wes06}. 

\begin{table*}
\begin{center}
\caption{Observing log of PSR~J1119$-$6127. The top portion of the table lists the fold-mode observations 
with the search-mode observations in the bottom portion. Measured spin frequency and flux density are 
presented for each observations. For search-mode observations, we also list the number of detection of strong 
single pulses (No. of SP) with $\rm{S/N}>10$. Profile types are as given in Section~\ref{sec:prof}.}
\label{num_pulse}
\begin{tabular}{lccccccccc}
\hline
\hline
\multicolumn{10}{c}{Fold mode} \\
\hline
	Date  &  MJD  &  Receiver   &   Frequency &  Bandwidth  &  Integration  & $\nu$  &   \multicolumn{2}{c}{Flux}   & Profile shape  \\
	      &       &             & (MHz)         &     (MHz)   & (s)           & (Hz)   & \multicolumn{2}{c}{(mJy)}    &                \\
\hline
	2016 May  22  & 57530.23  & HOH  & 1465   & 512    & 180     &  2.44000(2)     & \multicolumn{2}{c}{0.97(4)}     &  A   \\
	2016 Jun. 19  & 57558.18  & HOH  & 1465   & 512    & 180     &  2.43991(2)     & \multicolumn{2}{c}{1.01(4)}     &  A   \\
	2016 Jul. 24  & 57593.05  & HOH  & 1465   & 512    & 180     &  2.43986(1)     & \multicolumn{2}{c}{0.97(4)}     &  A   \\
	2016 Aug. 9   & 57609.14  & HOH  & 1369   & 256    & 970     &  2.439801(5)    & \multicolumn{2}{c}{0.29(1)}     &  A   \\
	2016 Aug. 9   & 57609.19  & HOH  & 1369   & 256    & 940     &  2.439811(6)    & \multicolumn{2}{c}{0.34(2)}     &  A   \\
	2016 Aug. 9   & 57609.33  & HOH  & 1369   & 256    & 600     &  2.439824(6)    & \multicolumn{2}{c}{0.56(2)}     &  A   \\ 
	2016 Aug. 11  & 57611.21  & HOH  & 1369   & 256    & 640     &  2.439803(6)    & \multicolumn{2}{c}{0.65(2)}     &  A   \\
	2016 Aug. 12  & 57612.03  & HOH  & 1369   & 256    & 460     &  2.439823(9)    & \multicolumn{2}{c}{1.05(3)}     &  B   \\
	2016 Aug. 12  & 57612.04  & HOH  & 1465   & 512    & 330     &  2.43981(1)     & \multicolumn{2}{c}{1.08(2)}     &  B   \\
	2016 Aug. 13  & 57613.37  & HOH  & 1369   & 256    & 500     &  2.439803(2)    & \multicolumn{2}{c}{2.18(3)}     &  B   \\
	2016 Aug. 15  & 57615.07  & HOH  & 1369   & 256    & 150     &  2.43977(2)     & \multicolumn{2}{c}{2.80(5)}     &  B   \\
	2016 Aug. 19  & 57619.12  & HOH  & 1369   & 256    & 300     &  2.43977(1)     & \multicolumn{2}{c}{3.77(5)}     &  B   \\
	2016 Aug. 19  & 57619.16  & HOH  & 1465   & 512    & 180     &  2.43979(1)     & \multicolumn{2}{c}{2.63(4)}     &  B   \\
	2016 Aug. 19  & 57619.18  & HOH  & 1369   & 256    & 780     &  2.439781(3)    & \multicolumn{2}{c}{3.08(3)}     &  B   \\
	2016 Aug. 19  & 57619.19  & HOH  & 1369   & 256    & 550     &  2.439780(4)    & \multicolumn{2}{c}{2.64(3)}     &  B   \\
	2016 Aug. 19  & 57619.20  & HOH  & 1369   & 256    & 350     &  2.43980(1)     & \multicolumn{2}{c}{2.41(4)}     &  B   \\
	2016 Aug. 19  & 57619.97  & HOH  & 1465   & 512    & 1200    &  2.439780(1)    & \multicolumn{2}{c}{2.69(2)}     &  B   \\
	2016 Aug. 24  & 57624.98  & HOH  & 1369   & 256    & 1200    &  2.439760(1)    & \multicolumn{2}{c}{3.15(2)}     &  B   \\
	2016 Aug. 25  & 57625.03  & HOH  & 1369   & 256    & 610     &  2.439756(4)    & \multicolumn{2}{c}{2.68(3)}     &  B   \\
	2016 Aug. 26  & 57626.02  & HOH  & 1369   & 256    & 490     &  2.439748(5)    & \multicolumn{2}{c}{2.55(3)}     &  B   \\
	2016 Aug. 29  & 57629.88  & HOH  & 1369   & 256    & 1200    &  2.43973(2)     & \multicolumn{2}{c}{3.31(3)}     &  C   \\
	2016 Aug. 29  & 57629.91  & HOH  & 1369   & 256    & 810     &  2.439724(1)    & \multicolumn{2}{c}{3.26(3)}     &  C   \\
	2016 Aug. 30  & 57630.09  & HOH  & 1369   & 256    & 1200    &  2.439723(1)    & \multicolumn{2}{c}{3.48(4)}     &  C   \\
	2016 Aug. 30  & 57630.17  & HOH  & 1369   & 256    & 1200    &  2.439725(1)    & \multicolumn{2}{c}{3.84(4)}     &  C   \\
	2016 Aug. 30  & 57630.21  & HOH  & 1369   & 256    & 1200    &  2.4397241(8)   & \multicolumn{2}{c}{4.29(4)}     &  C   \\
	2016 Aug. 31  & 57631.02  & HOH  & 1369   & 256    & 1200    &  2.4397157(8)   & \multicolumn{2}{c}{4.83(5)}     &  C   \\
	2016 Aug. 31  & 57631.03  & HOH  & 1369   & 256    & 1200    &  2.4397178(8)   & \multicolumn{2}{c}{5.05(5)}     &  C   \\
	2016 Aug. 31  & 57631.05  & HOH  & 1369   & 256    & 1200    &  2.4397178(8)   & \multicolumn{2}{c}{4.99(5)}     &  C   \\
	2016 Sep. 01  & 57632.14  & HOH  & 1369   & 256    & 520     &  2.439700(5)    & \multicolumn{2}{c}{4.35(8)}     &  C   \\
	2016 Sep. 01  & 57632.15  & HOH  & 1369   & 256    & 1200    &  2.4397054(9)   & \multicolumn{2}{c}{4.77(6)}     &  C   \\
	2016 Sep. 01  & 57632.17  & HOH  & 1369   & 256    & 1550    &  2.439706(1)    & \multicolumn{2}{c}{4.46(7)}     &  C   \\
	2016 Sep. 04  & 57635.87  & HOH  & 1369   & 256    & 1250    &  2.439670(1)    & \multicolumn{2}{c}{3.95(8)}     &  C   \\
	2016 Sep. 04  & 57635.91  & HOH  & 1369   & 256    & 400     &  2.439672(4)    & \multicolumn{2}{c}{3.83(4)}     &  C   \\
	2016 Sep. 25  & 57656.84  & HOH  & 1369   & 256    & 1380    &  2.439520(4)    & \multicolumn{2}{c}{0.93(7)}     &  A   \\
	2016 Oct. 14  & 57675.84  & HOH  & 1465   & 512    & 1200    &  2.439371(4)    & \multicolumn{2}{c}{0.29(1)}     &  A   \\
	2016 Nov. 12  & 57704.75  & HOH  & 1465   & 512    & 1200    &  2.439251(5)    & \multicolumn{2}{c}{0.14(1)}     &  A   \\
	2016 Dec. 10  & 57732.72  & MB   & 1369   & 256    & 1200    &  2.439186(4)    & \multicolumn{2}{c}{0.60(2)}     &  A   \\
	2017 Jan.  8  & 57761.60  & MB   & 1369   & 256    & 1200    &  2.43909(1)     & \multicolumn{2}{c}{0.31(2)}     &  A   \\
	2017 Feb. 11  & 57795.51  & MB   & 1369   & 256    & 1200    &  2.439046(6)    & \multicolumn{2}{c}{0.83(2)}     &  A   \\
	2017 Mar. 19  & 57831.45  & MB   & 1369   & 256    & 1200    &  2.438958(5)    & \multicolumn{2}{c}{0.82(2)}     &  A   \\
	2017 Apr. 09  & 57852.37  & MB   & 1369   & 256    & 1200    &  2.438925(4)    & \multicolumn{2}{c}{0.92(2)}     &  A   \\
	2017 Jun. 05  & 57909.21  & MB   & 1369   & 256    & 300     &  2.438803(4)    & \multicolumn{2}{c}{0.93(5)}     &  A   \\
\hline
\multicolumn{10}{c}{Search mode} \\
\hline
	Date   &    MJD        &    Receiver   &   Frequency &  Bandwidth     &   Integration  &   $\nu$    &  No. of SP &   Flux  & Pulse shape\\
	       &               &               &  (MHz)      &     (MHz)      &   (s)          &   (Hz)     &            &  (mJy)  &     \\
\hline
	2016 Aug. 9       & 57609.15  & HOH  & 1369   & 256    & 2586     &     2.439816(2)         &  0   &   0.21(1)   & A  \\
	2016 Aug. 9       & 57609.34  & HOH  & 1369   & 256    & 3355     &     2.439818(3)         &  0   &   0.21(1)   & A  \\
	2016 Aug. 12      & 57612.05  & HOH  & 1369   & 256    & 287      &     2.43981(1)          &  2   &   1.12(3)   & B  \\
	2016 Aug. 13      & 57613.38  & HOH  & 1369   & 256    & 1303     &     2.439805(1)         &  15  &   2.43(2)   & B  \\
	2016 Aug. 15      & 57615.08  & HOH  & 1369   & 256    & 363      &     2.439794(5)         &  17  &   3.15(3)   & B  \\
	2016 Aug. 29      & 57629.90  & HOH  & 1369   & 256    & 1123     &     2.4397243(9)        &  3   &   3.25(3)   & C  \\
	2016 Aug. 30      & 57630.19  & HOH  & 1369   & 256    & 1433     &     2.4397234(6)        &  8   &   3.07(2)   & C  \\
	2016 Sep. 1       & 57632.19  & HOH  & 1369   & 256    & 1866     &     2.4397051(2)        &  45  &   5.58(2)   & C  \\
	2016 Sep. 4       & 57635.89  & HOH  & 1369   & 256    & 1783     &     2.4396693(4)        &  1   &   3.22(2)   & C  \\
	2018 Apr. 20      & 58228.61  & MB   & 1369   & 256    & 1273     &     2.450800(3)         &  0   &   0.92(2)   & A  \\
\hline
\end{tabular}
\end{center}
\end{table*}

\section{Discussion}
\label{sec:discussion}

\begin{figure}
\begin{center}
\includegraphics[width=3.2 in]{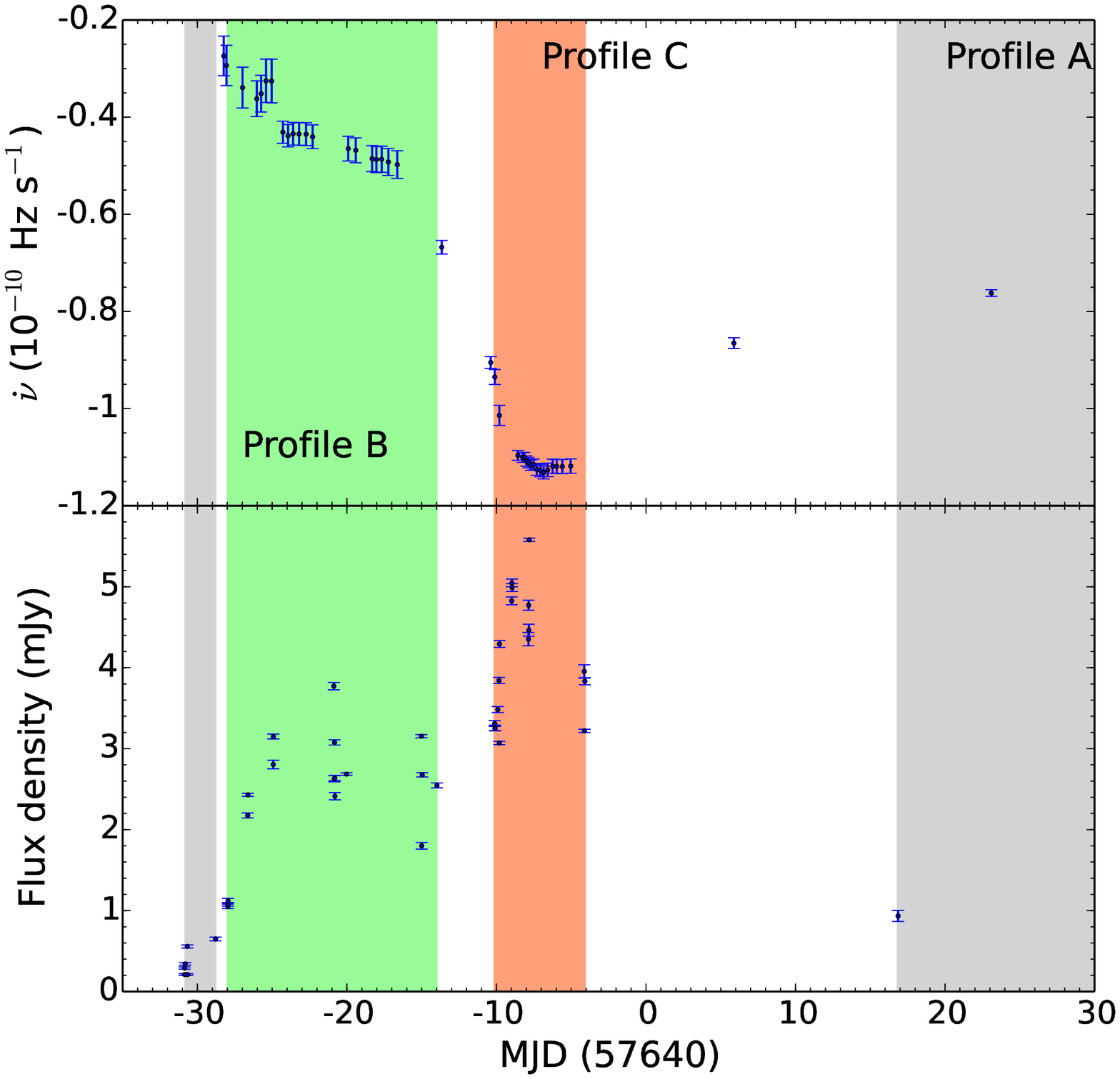}
\caption{Top: $\dot\nu$ as a function of time; Bottom: flux density as a 
function of time. Grey, green and red shaded regions represent pulse profiles 
in type A, B and C, respectively. We fitted for $\dot\nu$ by using 
adjacent measurements of $\nu$ with a sliding window as described in Section~\ref{sec:nu}, 
and therefore there are $\dot\nu$ points in between different pulse profile types.}
\label{f0_zoom}
\end{center}
\end{figure}

We observe a striking correlation between the profile changes, the flux density 
variability and the changes in the timing, particularly the value of $\dot{\nu}$.
In Fig~\ref{f0_zoom}, we show the evolution of $\dot\nu$ and flux density 
as a function of time directly following the X-ray bursts within 65 days. 
Different pulse profile types are indicated with shaded regions.
During the rapid evolution of spin frequency (within the first 30 days), the pulse 
profile changed twice, on 2016 August 12 ($\rm{MJD}\approx57612$)
and 29 ($\rm{MJD}\approx57630$), and at both times we observed fast increase of 
flux density and decrease of $\dot{\nu}$ (becomes more negative). We also see that 
for Profile B and Profile C the $\dot\nu$ is significantly different by $\sim0.6\times10^{-10}$\,Hz\,s$^{-1}$.
However, the pulsar has then reverted to its normal profile ($\rm{MJD}\approx57656$) 
prior to the slow recovery of $\dot{\nu}$ to its original value.

It therefore seems likely that the torque acting on the pulsar causing the
change in $\nu$ is also related to the changing pulse profile.
This is seen in other pulsars, more specifically in the mode changing
pulsars. \citet{lhk+10} noticed that pulse profile changes were correlated with
$\dot{\nu}$ changes. In the pulsars they examined, these changes occurred
quickly but then persisted for a timescale of months~\citep[see also][]{ksj13,bkj+16}. 
In the intermittent pulsar B1931$+$24~\citep{klo+06} the $\dot{\nu}$ changes
appear to be caused by plasma loading in the pulsar magnetosphere.
We surmise that something similar is happening in this case.

The only other example of a high-$B$ pulsar transitioning to a magnetar observed so far is 
the 2006 magnetar-like outburst of PSR J1846$-$0258~\citep{ggg+08}. A large but typical 
glitch recovery has been detected for J1846$-$0258 after the glitch triggered by the X-ray 
bursts~\citep{lkg10}.
However, after the reactivation of radio radiation of PSR J1119$-$6127, we did not observe 
significant discrete changes of $\nu$ and $\dot\nu$. On the contrary, a smooth and fast spin-down 
process with a net spin-down rate change of $\sim10^{-10}$\,Hz\,s$^{-1}$ has been observed, 
which is much larger than that of PSR J1846$-$0258 ($\sim8\times10^{-12}$\,Hz\,s$^{-1}$)
and is clearly different from normal glitch recoveries.
\citet{crp+18} present the results of a systematic study of all magnetar 
X-ray outbursts observed to date and show that the total outburst energy of 
J1119$-$6127 ($8.5\times10^{41}$\,erg) is larger than that of 
J1846$-$0258 ($4.5\times10^{41}$\,erg). On the other hand, as discussed by \citet{akt+16}, 
the X-ray spectra of J1119$-$6127 in quiescence and during the outburst show different 
spectral shape and evolution compared with that of J1846$-$0258.
The differences between the behaviour of J1119$-$6127 and J1846$-$0258 suggest 
that the peculiar spin frequency evolution of J1119$-$6127 reveal processes not well 
understood during the transition from a high-$B$ radio pulsar to a magnetar.

The net spin-down we observed following the X-ray bursts implies a release of a large amount of 
rotational energy. Assuming the canonical neutron star moment of inertia of $I_{45}=10^{45}$\,g\,cm$^2$, 
the rotational energy released can be estimated as
\begin{equation}
	\Delta E_{\rm{rot}}=4\pi^{2}\nu|\Delta\nu|=3.8\times10^{43}I_{45}\left(\frac{|\Delta\nu|}{4\times10^{-4}}\right)\ \rm{erg}.
\end{equation}
This is $\sim20$ times of the total X-ray outburst energy, and in fact we did not 
observe obvious correlation between the X-ray light curve presented in \citet{crp+18} and
the radio flux densities nor the spin-frequency evolution.
For a time span of $\sim100$\,days, the spin-down power gives us an averaged luminosity 
of $\sim4.4\times10^{36}$\,erg\,s$^{-1}$, about two times larger than the normal 
spin-down luminosity.
While only a small portion of the spin-down energy goes into radio and X-ray, we could 
expect much stronger high energy emissions from the pulsar (e.g., $\gamma$-ray) if energy 
loss via high energy radiation is the main energy-loss mechanism. Such a high spin-down 
luminosity could also power strong relativistic particle outflows that can brighten the 
pulsar wind nebula~\citep{bsm17}.

A complete picture of the PSR J1119$-$6127 event is likely to be that ultra-strong magnetic 
fields powered the X-ray bursts and triggered an additional torque that produced
the peculiar spin frequency evolution.
The fact that the polarization position angle changed significantly following the X-ray bursts 
indicates that the magnetic field might have changed.
If we assume that the spin-down process is dominated by dipole braking, the surface 
magnetic field strength can be estimated as
\begin{equation}
	\label{field}
	B_{\rm{S}}\approx10^{12}\,\rm{G}\left(\frac{\dot{P}}{10^{-15}}\right)^{1/2}\left(\frac{P}{\rm{s}}\right)^{1/2},
\end{equation}
where $P$ and $\dot{P}$ are the spin period and spin-period derivative.
Applying Eq.~\ref{field} on our measurements of $P$ and $\dot{P}$ of PSR J1119$-$6127, we 
found that after the X-ray burst the surface magnetic field strength can reach 
as high as $\sim9\times10^{13}$\,G, which is more than two times stronger than 
the normal field strength. 
Under such assumptions, the fast increase and slow recovery of $\dot\nu$ as shown 
in the third panel of Fig.~\ref{f0} indicate similar evolution of the surface magnetic 
field strength of the pulsar. This suggests that the X-ray bursts triggered some
reconfiguration of the magnetosphere of the pulsar, resulting in the increase of 
the apparent field strength. 
The energy of the global dipolar field external to the neutron star can be estimated as
\begin{equation}
	E_{\rm{dipole}} = \frac{10^{42}}{3}\left(\frac{B_{\rm{S}}}{10^{12}\ \rm{G}}\right)^{2}\left(\frac{R}{10\ \rm{km}}\right)^{3}\ \rm{erg},
\end{equation}
and in order to increase the magnetic field strength from $\sim4\times10^{13}$\,G to 
$\sim9\times10^{13}$\,G a total amount of energy of $\sim2\times10^{45}$\,erg is needed. 
This is more than one order of magnitude larger than the spin-down power released 
following the X-ray bursts. Energies either from multipole (and/or internal) magnetic 
fields or neutron star interior need to be converted to the strong dipolar field and/or 
distort the magnetic field. Under such a scenario, the slow recoveries of the 
timing behaviour and the pulse profile back to normal have to be explained, 
which might involve the dissipation of the magnetic field.

As an alternative solution wind braking models have been suggested to explain the spin-down of magnetars~\citep[e.g.,][]{hck99,txs+13} 
and the anti-glitch of magnetar 1E 2259$+$586~\citep{tong14}. Particle winds triggered by the 
X-ray burst~\citep[e.g.,][]{td96} can amplify the rotational energy loss rate
by providing additional torque to slow down the pulsar.
Enhanced particle winds could also be responsible for the fast increase of radio flux
density and dramatic changes of pulse profile.
Generalizing the model \citet{klo+06} proposed to understand different spin-down rates when  
the intermittent pulsar B1931$+$24 is on and off, we can estimate  
the change in charge density in the open field line region as 
\begin{equation}
\label{rho}
	\frac{\Delta\rho}{\rho_{\rm{GJ}}}=-2.0\times\left(\frac{R}{10^{4}\ \rm{m}}\right)^{-6}\left(\frac{B_{\rm{s}}}{10^{8}\ \rm{T}}\right)^{-2}\left(\frac{\nu}{1\ \rm{Hz}}\right)^{-3}\left(\frac{\Delta\dot{\nu}}{10^{-15}\ \rm{Hz}\ \rm{s}^{-1}}\right),
\end{equation}
where $R$ is the radius of a neutron star and $\rho_{\rm{GJ}}$ is the Goldreich-Julian density~\citep{gj69}.
At the maximum net spin-down rate ($\Delta\dot{\nu}\approx10^{-10}$\,Hz\,s$^{-1}$), 
Eq.~\ref{rho} gives us a net change in charge density of $\Delta\rho/\rho_{\rm{GJ}}\approx8.4$, 
which means that the change in charge density in the open field line region needs to be much 
larger than $\rho_{\rm{GJ}}$.
Compared with XTE J1810$-$197, whose decrease of spin-down rate and flux density
and changes of pulse profile have been argued to be caused by a net change of plasma 
density of $\Delta\rho/\rho_{\rm{GJ}}\approx0.2$~\citep{ccr+07}, the extra plasma 
we need to explain the torque of PSR J1119$-$6129 is much larger.
It is not clear if such high magnetospheric charge densities can be sustained long 
enough to explain the spin-down of J1119$-$6127.
More realistic pulsar wind and particle acceleration models need to be taken into 
consideration in order to obtain self-consistent explanations~\citep[e.g.,][]{lty+14,xq01}.

\section{Summary}
\label{sec:sum}
We presented the spin frequency, flux density and pulse profile evolution of PSR~J1119$-$6127
following the X-ray bursts detected in July, 2016, using data taken with the Parkes 
radio telescope. While PSR~J1119$-$6127 showed radio properties, such as dramatic changes 
of pulse profile and flux density variability, shared by other magnetars, the 
peculiar spin frequency evolution is clearly different from what has been observed from 
magnetars and high-$B$ pulsars before. Therefore, as the first rotation-powered, 
radio pulsar to show magnetar-like activities, PSR~J1119$-$6127 provides us the best 
case to study the connection between high-B pulsars and magnetars.


\section*{Acknowledgements}

The Parkes radio telescope is part of the Australia Telescope
National Facility which is funded by the Commonwealth of
Australia for operation as a National Facility managed by CSIRO.
Work at NRL is supported by NASA.
P.E. and N.R. acknowledge funding in the framework of the NWO Vidi award A.2320.0076.


\bibliography{ms}


\bsp	
\label{lastpage}
\end{document}